\documentclass[%
reprint,
showpacs,
showkeys,
amsmath,amssymb,
aps,
prd,
]{revtex4-2} 

\usepackage{graphicx}
\usepackage{pgffor}
\usepackage{capt-of}
\usepackage{soul}
\usepackage{lipsum}

\newcommand{\uvozovky}[1]{``#1''}

\def\beq{\begin{equation}}
\def\eeq{\end{equation}}
\def\bea{\begin{eqnarray}}
\def\eea{\end{eqnarray}}

\let\phi=\varphi

\let\phi=\varphi
\let\rho=\varrho

\begin{document}

\frenchspacing

\author{Martin Blaschke}
\email{martin.blaschke@physics.slu.cz}
\author{Zden\v{e}k Stuchl\'{\i}k}
\email{zdenek.stuchlik@physics.slu.cz}
\author{Sudipta Hensh}
\email{F170656@physics.slu.cz,sudiptahensh2009@gmail.com}
\affiliation{%
Research Centre for Theoretical Physics and Astrophysics, Institute of Physics, Silesian University in Opava,\\
Bezru\v{c}ovo n\'am\v est\'i~13, CZ-746\,01 Opava, Czech Republic%
}

\title[Evolution]{Evolution of Braneworld Kerr--Newman Naked Singularities}

\begin{abstract}
We study evolution of the braneworld Kerr--Newman (K-N) naked singularities, namely their mass $M$, spin $a$, and tidal charge $b$ characterizing the role of the bulk space, due to matter in-falling from Keplerian accretion disk. We construct the evolution in two limiting cases applied to the tidal charge. In the first case we assume $b=$ const during the evolution, in the second one we assume that the dimensionless tidal charge $\beta \equiv b/M^2 =$ const. For positive values of the tidal charge the evolution is equivalent to the case of the standard K-N naked singularity under accretion of electrically neutral matter. We demonstrate that counter-rotating accretion always converts a K-N naked singularity into an extreme K-N black hole and that the corotating accretion leads to variety of outcomes. The conversion to an extreme K-N black hole is possible for naked singularity with dimensionless tidal charge $\beta<0.25$, and $\beta\in(0.25,1)$ with sufficiently low spin. In other cases the accretion ends in a transcendental state. For $0.25<\beta<1$ this is a mining unstable K-N naked singularity enabling formally unlimited energy extraction from the naked singularity. In the case of $\beta>1$, the corotating accretion creates unlimited torodial structure of mater orbiting the naked singularity. Both non-standard outcomes of the corotating accretion imply a transcendence of such naked singularity due to nonlinear gravitational effects.     
\end{abstract}


\maketitle 
\section{Introduction}

The discovery of exact solutions to Einstein field equations describing rotating black holes (Kerr and Kerr--Newman) had influence on practically every subfield of general relativity and revolutionized astrophysics \cite{Teu:2015:}.  Kerr solution \cite{Ker:1963:} is mainly important in the astrophysics of quasars and accreting stellar-mass black hole systems. More general solution containing electric charge, the K-N solution \cite{New:etal:1965:}, is widely regarded as unimportant to astrophysical phenomena, because any net charge should be quickly neutralized due to presence of plasma. However, we still can assume situations allowing, at least for a limited time, for presence of a small electric charge in a rotating K-N black hole influencing astrophysical processes \cite{Ruf:1973:BlaHol:,Stu-etal:2021:submitted:}. Of special interest could be the K-N spacetimes containing (hypothetical) magnetic charge \cite{Cal:Stu:1983:,Stu:1983:BAC:}, as the magnetic charge cannot be neutralized by accretion of electric charge so it is surely relevant to consider also the K-N backgrounds. Moreover, the external large-scale magnetic fields around rotating black holes can induce an electric charge in the black holes having extended astrophysical consequences as demonstrated in series of papers \cite{Kol-Stu-Tur:2015:CLAQG:,Stu-Kol:2016:EPJC:,Tur-Stu-Kol:2016:PHYSR4:,Kol-Tur-Stu:2017:EPJC:,Tur:etal:2020:,Stu:etal:2020:,Kol-etal:2021:PHYSR4:}.  

All the black hole solutions admit extension into the regime with no horizons, i.e. naked singularity (NS) spacetimes. 
The possible existence of NS's in the nature is still interesting and unresolved issue, because the Penrose cosmic censorship conjecture \cite{Pen:1969:,Haw:Ell:1973:} has not been proven within the framework of GR yet (see e.g. \cite{Jos:1993:,Pen:1972:,Jos:Mal:2011:}). Therefore, it is interesting to study how would NS's influence astrophysical relevant phenomena and compare them to their black hole counterparts (see, e.g. \cite{Gos:Jos:2007:}). 

Kerr naked singularities demonstrate some astrophysical phenomena that could be considered extraordinary even in comparison to the related black-hole phenomena in both accretion \cite{Stu:1980:BAC:,Stu:Hle:Tru:2011:,Stu-Sche:2012:CLAQG:,Stu-Sche:2013:CLAQG:} and optical effects \cite{Stu:Hle:2000:,Stu-Sche:2010:CLAQG:,Char-Stu:2017:EPJC:,Stu-Char-Sche:2018:EPJC:,Tro-etal:2018:}. One possibility can be to try to find general differences in astrophysical processes such as accretion and lensing. In the case of spherically symmetric naked singularities many studies were presented; for example  with regards to accretion in \cite{Kov-Har:2010:PRD:,Jos:Mal:Nar:2014:,Bic-Stu-Bal:1989:BAC:,Bal-Bic-Stu:1989:BAC:,Stu-Bic-Bal:1999:GRG:}, and for strong lensing in \cite{Vir:atal:1998:,Vir:Ell:2002:}.

The braneworld version of the K-N solution, which describes axially symmetric rotating black hole localized on the 3Dbrane in Randall--Sundrum II model, can be also astrophysically very important giving imprints of hidden dimensions. The tidal charge $b$ representing a non-local imprint of the additional bulk dimension in the braneworld models, is only the analog of electric charge in the K-N spacetime. This object is actually electrically neutral \cite{Ali-Gum:2005:CLAQG:,Dad:2000:}, having no electromagnetic field. The astrophysical consequences of the existence of the braneworld black holes were extensively studied in the literature \cite{Kot-Stu-Tor:2008:CLAQG:,Sche:Stu:2009b:,Sche:Stu:2009:,Stu-Kot:2009:GRG:,Ali-Tal:2009:prd:,Nun:2010:PHYSR4:,Ami-Eir:2012:PHYSR4:}. The case of related K-N naked singularities was studied in relation to the Keplerian accretion disks in \cite{Bla:Stu:2016:}. 

In the case of classical and also braneworld K-N spacetime, the extensive classification with respect to Keplerian accretion properties was done in \cite{Bla:Stu:2016:,Stu:Bla:Sche:2017:}. A strange new phenomenon, so called mining instability, was discovered for NS class IIIa. This instability could in principle solve problems of very high energy particles present in the Universe (see also \cite{Stu:etal:2020:,Tur:etal:2020:,Kol-etal:2021:PHYSR4:}).

Here we study the evolution of the K-N naked singularities due to the simple model of Keplerian accretion, extending thus the results of the evolution of Kerr naked singularities presented in \cite{Stu:1981:BULAI:,Stu:Hle:Tru:2011:}. The basic question is, if all K-N naked singularities could be converted into black holes due to Keplerian accretion. Recall that standard accretion disks cannot exist in the K-N naked singularity mining unstable spacetimes. Furthermore, we show that some classes of K-N naked singularities (with $b>M^2$) capture the accreting mass in the Polish donut like structures similar to black Saturn solution. This means that matter cannot fall into the singularity, but the accreting matter, considered under test particle motion approximation, is instead gathered in the minimum of the effective potential of the test particle motion applied in the Keplerian accretion model.       

We also show that some classes of braneword K-N black holes with negative values of tidal charge ($b\equiv Q^2<0$) do not evolve due to accretion to their extremal cases, as it is usual, but instead they remain unchanged. They are \uvozovky{frozen} braneworld black holes with respect to corotating accretion. We have to stress that the counterrotating accretion regime can overcome such \uvozovky{frozen} states.          

\section{Spacetime geometry}\label{models}

We study a simple model of Keplerian thin accretion disk and consequently its influence on the spacetime evolution due to the accreting matter falling onto the central object. We consider the central object to be naked singularity described via generalized K-N metric - so-called braneworld K-N metric, where the charge parameter could be also negative (for more details and derivation of this metric see \cite{Ali-Gum:2005:CLAQG:}).

In the Boyer--Lindquist coordinates $(t,r,\theta,\varphi)$ and the geometric units $(c=G=1)$, the~line element of the braneworld K-N metric, representing solution of the Einstein equations induced on the 3D-brane, reads \cite{Ali-Gum:2005:CLAQG:,Dad:2000:} 

\begin{eqnarray}
\mathrm{d}s^2= &-& \left(1-\frac{2Mr - b}{\Sigma }\right)\mathrm{d}t^2 + \frac{\Sigma}{\Delta}\,\mathrm{d}r^2 - \nonumber \\
& &\frac{2a(2Mr - b)}{\Sigma}\,\mathrm{sin}^2\theta\,\mathrm{d}t\mathrm{d}\phi + \Sigma\, \mathrm{d}\theta^2 +\, \nonumber \\
& & \left(r^2 + a^2 + \frac{2Mr - b}
{\Sigma}\,a^2\mathrm{sin}^2\theta\right)
\mathrm{sin}^2\theta\, \mathrm{d}\phi^2\, , \label{Metrika}
\end{eqnarray}
with 
\begin{eqnarray}
\Delta &=& r^2 -2Mr +a^2 + b\, ,\\ 
\Sigma &=& r^2 + a^2\mathrm{cos}^2\theta\, ,
\end{eqnarray}
where $M$ is the mass parameter of the spacetime, $a=J/M$ is the~specific angular momentum (spin) of the spacetime with internal angular momentum $J$, and the~braneworld tidal charge parameter $b$ represents imprint of the non-local gravitational effects of the~bulk space \cite{Ali-Gum:2005:CLAQG:,Ran:Sun:1999b:}. 

The Keplerian model of thin accretion disk is useful to study evolution of the central object due to its simplicity. It has been shown that a BH never evolves via Keplerian accretion into a NS (see, e.g. \cite{Bar:1970:,Page:1976:}). A Kerr naked singularity evolution due to the accretion process will always eventually end up approaching an extremal BH \cite{Cal:Nob:1979:,Stu:1981:BULAI:}. 

In this work we stress the special role of newly discovered class of K-N naked singularity spacetimes which suffers so-called mining instability \cite{Bla:Stu:2016:}. This class (IIIa due to \cite{Bla:Stu:2016:}) was originally studied under 5D branewold paradigm, but the results are valid even for classical four-dimensional (4D) case, because of the simple relationship with its braneworld equivalent. The 4D induced metric on the brane can be formally obtained from the K-N metric by substitution $Q^2\rightarrow b\, ,$ where $Q$ is the electric charge of the spacetime \cite{Shi:Mae:Sas:2000:}.

\section{Evolutionary equations of spacetime parameters}

In order to follow evolution of the K-N naked singularity due to Keplerian accretion, i.e., accretion of matter from the marginally stable orbit, we are using completely dimensionless quantities describing particular spacetime: 
\begin{equation}
\alpha = \frac{a}{M}\, , \quad \beta = \frac{b}{M^2}\, .
\end{equation}

As we have no hints on the reaction of the bulk space due to the tidal effects affected by the accretion process and related modifications of the parameters describing the central naked singularity, we have to use simple rules limiting possible changes of the tidal charge during evolution. 

Therefore, in the following we will focus on two models of Keplerian accretion as related to the assumed change of parameter $b$ during accretion:
\begin{enumerate}
\item Model where we consider the dimensionless tidal charge $\beta$ being constant during accretion process.   
\item Model where the tidal charge $b$ is constant during accretion. This model has one advantage, namely, for positive tidal charge ($b>0$) it is equivalent to the case of accretion of non-charged matter onto standard K-N BH or NS. 
\end{enumerate}

Note that these two choices  represent in a sense limiting cases given on the behaviour of evolution of the tidal charge during accretion of matter. However, the case of $\beta =$ const for $\beta>1$ excludes final transformation of a naked singularity into a black hole. 
  
For both these models of thin Keplerian accretion, the infinitesimal change of the mass $dM$ and angular momentum $dJ$ is given by relations:
\begin{eqnarray}
d M &=& E_{\mathrm{ms}}\,  d\mu\, ,\label{mem} \\
d J &=& L_{\mathrm{ms}}\,  d\mu\, ,
\end{eqnarray}
where $E_{\mathrm{ms}}$ is the specific energy and $L_{\mathrm{ms}}$ is the specific angular momentum of the particle with rest mass $\mu$ which is located at the edge of the accretion disk - at marginally stable (ms) orbit.

We can directly follow \cite{Tho:1974:} and find the master equation governing evolution of the central body due to accretion in the form: 
\begin{equation}\label{dr}
\frac{d \alpha}{dM}= \frac{1}{M^2}\frac{L_{\mathrm{ms}}}{E_{\mathrm{ms}}}-\frac{2\alpha}{M}\, .
\end{equation}

In the K-N spacetime, the specific energy $E(x,\alpha,\beta)$ and specific angular momentum $L(x,\alpha,\beta)$ of a test particle following equatorial circular geodesic at dimensionless radius $x=r/M$ are given by relations \cite{Stu:1980:BAC:,Ali-Gum:2005:CLAQG:,Dad-Kal:1977:,Stu-Kot:2009:GRG:}:
\begin{eqnarray}
E(x,\alpha,\beta) &=& \frac{x^2-2 x +\beta \pm \alpha \gamma}{x\sqrt{x^2-3x  +2\beta\pm 2\alpha\gamma}}\, ,\label{E} \\
L(x,\alpha,\beta) &=& \pm M \frac{\gamma\left(x^2+\alpha^2 \mp 2\alpha\gamma\right)\mp \alpha\beta}{x\sqrt{x^2-3x +2\beta\pm 2\alpha\gamma}}\, ,\label{L} 
\end{eqnarray}
where
\begin{equation}
\gamma = \sqrt{x-\beta}\, .
\end{equation}

The + sign corresponds to upper family, - sign to the lower family of circular orbits. We use prefix \uvozovky{upper} and \uvozovky{lower} because the usual distinction between co-rotating  and counter-rotating orbits are in the case of spacetime with NS more complicated and cannot be resolved just by the $\pm$ sign. However, for simplicity sake, it is convenient to consider prefix \uvozovky{upper} resp. \uvozovky{lower} as almost synonymous to \uvozovky{co-rotating} resp. \uvozovky{counter-rotating}. The \uvozovky{lower} family geodesic circular orbits are always counter-rotating relative to distant observers being at any allowed radius $r\, .$ The \uvozovky{upper} family circular geodesics are always corotating with respect to distant observers at large radii, but could become counter-rotating in relation to distant observers around K-N naked singularities with spin close to the near-extreme value, and in limited region of radii around $r=M$ \cite{Stu:1980:BAC:}. 

The inner edge of the Keplerian accretion disk is located at (dimensionless) radii of the marginally stable circular orbits $x_{\mathrm{ms}}$, implicitly given by the condition \cite{Stu-Kot:2009:GRG:,Ali-Gal:1981:}
\begin{eqnarray}\label{rms}
\hspace{-5mm}
x(6x - x^2 - 9\beta +3\alpha^2) + 4\beta(\beta-\alpha^2)\mp 8 \alpha\gamma^{3}=0.
\end{eqnarray}
It is difficult to explicitly solve this equation with respect to $x$, but with respect to $\alpha$ it is just a quadratic equation the solution of which can be given in the form  
\begin{equation}\label{amsp}
\alpha_{\mathrm{ms}}=\pm\frac{4 \gamma^{3}\pm x_{\mathrm{ms}} Y(x_\mathrm{ms},\beta)}{\left(3x_{\mathrm{ms}}-4\beta\right)}\, ,
\end{equation}
where we have used abbreviation
\begin{equation}
Y(x_\mathrm{ms},\beta)= \sqrt{3 x_{\mathrm{ms}}^2 - 4x_{\mathrm{ms}} \beta-2x_{\mathrm{ms}} + 3 \beta}\, .
\end{equation}
The first $\pm$ sign in Eq. (\ref{amsp}) corresponds to two families of orbits and second $\pm$ to two possible solutions of the quadratic equation. As we are considering $\alpha>0\, ,$ we take the first ($+$ sign) family orbits with solutions of both signs, while the second ($-$ sign) family orbits we have to consider with the only  ($+$ sign) form, as the other one gives only negative values of spin. 



In the following sections we consider only the upper family orbits and deal with lower family orbits in separated section. 

\section{Future-oriented motion}

For the positive-root states (see \cite{Gravitation, Bic-Stu-Bal:1989:BAC:,Bla:Stu:2016:}) the time evolution has to be oriented to the future, i.e., $dt/d\tau >0$. During our analysis we could in principle mix positive-root states with (unphysical in our study) negative-root states. Therefore it is necessary to check that the considered geodesics have proper orientation $dt/d\tau >0$.

In general stationary and axially symmetric spacetime endowed with the Boyer--Lindquist coordinate system, the geodesic equation of the equatorial motion takes the form 
\begin{equation}\label{48}
\frac{dt}{d\tau} = \frac{E g_{\phi\phi}+L g_{t\phi}}{g^2_{t\phi}-g_{tt}g_{\phi\phi}}\, .
\end{equation}
K-N equatorial circular motion can be found using the metric (\ref{Metrika}) and relations for the specific energy (\ref{E}) and specific angular momentum (\ref{L}). We find that the sign of $dt/d\tau$ is governed by the sign of the relation
\begin{equation}
T\equiv \frac{x^2\pm\alpha \sqrt{x-\beta}}{x\sqrt{x^2-3x  +2\beta\pm 2\alpha\gamma}}\, ,
\end{equation}
where $\pm$ distinguish upper (+) and lower (-) family of orbits. 

A simple analysis of the expression $T$ demonstrates that for the case considered here, we are fixed only to the positive-root states with $T>0\, .$ 

\section{K-N spacetime classification}

\begin{figure}[t]
	\begin{center}
		\centering
		\includegraphics[width=\linewidth]{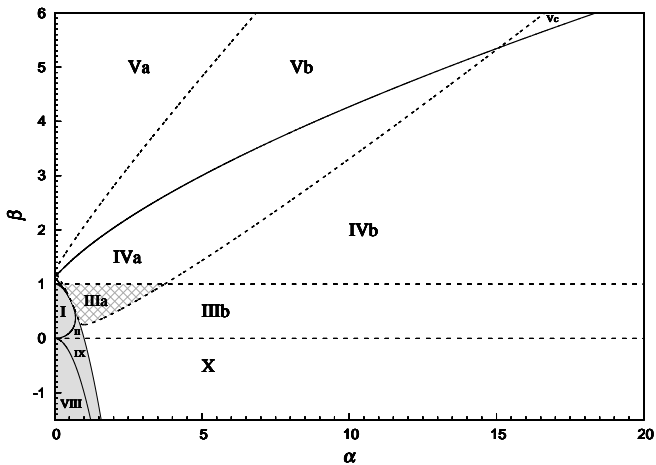}
		\caption{\label{regionab2} Classification of the~braneworld K-N spacetimes according to the~properties of circular geodesics relevant for the~Keplerian accretion. The parameter space $\beta-\alpha$ is separated by curves governing the~extrema of the~functions determining the~photon circular orbits (thick lines) and the~marginally stable orbits (dashed lines).}   
	\end{center}
\end{figure}

\begin{figure}[t]
	\begin{center}
		\centering
		\includegraphics[width=\linewidth]{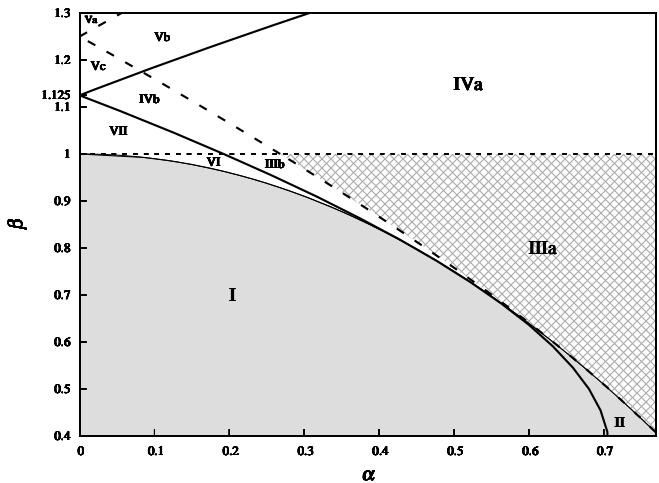}
		\caption{\label{regionab3} Classification of the~braneworld K-N spacetimes according to the~properties of circular geodesics relevant for the~Keplerian accretion. The parameter space $\beta-\alpha$ is separated by curves governing the~extrema of the functions determining the~photon circular orbits (thick lines) and the~marginally stable orbits (dashed lines). Detailed structure for small values of spin $\alpha$ and $\beta \sim 1$.}   
	\end{center}
\end{figure}

In order to understand the evolution of the braneworld K-N naked singularities, we have to apply the classification of the braneworld K-N spacetimes in relation to the Keplerian accretion as introduced in \cite{Bla:Stu:2016:}. We have to stress that the standard Keplerian accretion with matter freely falling onto the central object (NS or BH) from the marginally stable circular orbit is not allowed in some classes of the K-N naked singularity spacetimes where the accretion has to imply final occurrence of some transcendental states.   

Classification of the K-N (and also K-N braneworld) spacetimes with respect to Keplerian accretion properties was done in \cite{Bla:Stu:2016:} and \cite{Stu:Bla:Sche:2017:}. The classification is summarized in the Table \ref{class} and is reflected by Figs. \ref{regionab2} and \ref{regionab3}. The parameters $\alpha$ and $\beta$ separate the K-N spacetimes into 14 classes with respect to properties like location of the marginally stable orbits and number of photon orbits. In this article we focus on classes surrounding class IIIa (the mining instability class) which has the most interesting properties regarding accretion disks. Namely, we focus on classes I, II, IIIb, IVa, VI.

Note that for K-N naked singularities with initial value of spin $\alpha$ higher than that corresponding to the IIIa class of mining unstable naked singularities, the evolution usually ends in the IIIa state and in this state it cannot evolve to the extreme black hole state, but rather undergoes transcendence to some different state (black hole cannot be excluded) tractable only in fully non-linear regime of GR or braneworld model. The cause is the special \uvozovky{mining} character of the effective potential (see \cite{Bla:Stu:2016:}). In Fig. \ref{regionab2} we give relevant regions of the $\alpha-\beta$ space as related to possible accretion states. The extraordinary states of IIIa class are crucial for the evolution process.       

\begin{table}[t]
\begin{center}
\begin{tabular}{|c|c|c|c|c|c|c|}
\hline
\bfseries Class &
\bfseries ISCO &
\bfseries MSO(u) &
\bfseries MSO(l) &
\bfseries Hor./Erg. &
\bfseries SP &
\bfseries UP \\
\hline
I & =MSO & classic & classic & yes/yes & 0 & 2 \\ \hline
II & =MSO & classic & classic & yes/yes & 1 & 3 \\ \hline
IIIa & =Photon & $\displaystyle -$ & classic & no/yes & 1 & 1 \\ \hline
IIIb & =MSO & classic & classic & no/yes & 1 & 1 \\ \hline
IVa & at $x=\beta$ & $\displaystyle -$ & classic & no/no & 1 & 1 \\ \hline
IVb & at $x=\beta$ & classic & classic & no/no & 1 & 1 \\ \hline
Va & at $x=\beta$ & $\displaystyle -$ & $\displaystyle -$ & no/no & 0 & 0 \\ \hline
Vb & at $x=\beta$ & $\displaystyle -$ & classic & no/no & 0 & 0 \\ \hline
Vc & at $x=\beta$ & classic & classic & no/no & 0 & 0 \\ \hline
VI & =MSO & classic & classic & no/yes & 2 & 2 \\ \hline
VII & at $x=\beta$ & classic & classic & no/no & 2 & 2 \\ \hline
VIII & =MSO & classic & classic & yes/yes & 0 & 2 \\ \hline
IX & =MSO & classic & classic & yes/yes & 0 & 3 \\ \hline
X & =MSO & classic & classic & no/yes & 0 & 1 \\ \hline
\end{tabular}
\caption{\label{class} Classification of the K-N spacetime in the parameter space $\beta-\alpha$, with respect to: ISCO - radius of innermost stable circular orbit; MSO(u) - radius of Marginally Stable Orbit for upper sign family; MSO(l) - radius of Marginally Stable Orbit for lower sigh family; SP - number of stable photon circular orbits; UP - number of unstable photon circular orbits. ISCO has only two possible outcomes. It can be identical with MSO or lies at $x=\beta$. Word \uvozovky{classic} in this context means that MSO is defined by equation (\ref{rms}). Hor. - existence of the horizon (black hole), Erg. - existence of the ergosphere.} 
\end{center}
\end{table} 
   
\section{Energy sign of marginally stable circular orbits}

\begin{figure}[t]
	\begin{center}
		\includegraphics[width=1\linewidth]{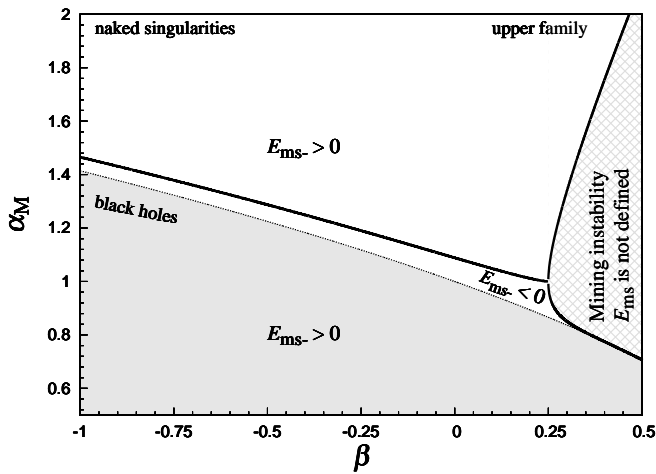}
	\end{center}
	\caption{\label{am} Functions $\alpha_{\mathrm{M}}(\beta)$ (black line) and $\alpha_{\mathrm{h}}$ (border of black holes region) governing the change of sign for $E_{\mathrm{msu}}$. Event horizon represents discontinuous jump in energy. The mining instability region is defined by condition $2\sqrt{\beta}-\sqrt{\beta(4\beta-1)}<\alpha  <2\sqrt{\beta}+\sqrt{\beta(4\beta-1)}\, $. For more details see \cite{Bla:Stu:2016:,Stu:Bla:Sche:2017:,Sha:2019:,Pug:Q:2018:}. }
\end{figure}

To study the influence of accretion on the central body, it is important to know the sign of the marginally stable circular orbit energy $E_{\mathrm{ms}}\equiv E(x_\mathrm{ms},\alpha,\beta)$. We see that this sign solely determines whether the central body mass grows or shrinks due to Keplerian accretion (see Eqs. (\ref{mem}) and (\ref{eM})). 

The energy $E_{\mathrm{ms}}=0$ can be reached by properly chosen K-N naked singularities for the upper family \uvozovky{corotating} Keplerian disks; if the extreme K-N state is reached, there is an significant jump of the energy, discussed in details in \cite{Stu:1980:BAC:}. On the other hand, counter-rotating Keplerian disks have for all the K-N naked singularity spacetimes $E_{\mathrm{ms}}>0$ and this energy is smoothly matched to the energy state $E_{\mathrm{ms}}$ corresponding to marginally stable counter-rotating orbit of the extreme K-N black hole.      

By combining and simplifying Eqs. (\ref{E}), (\ref{L}) and (\ref{amsp}), we find the relations for energy and angular momentum at marginally stable circular orbit. The evolution of a K-N naked singularity due to corotating Keplerian accretion towards the extreme black hole (or some transcendental) state is thus governed by the relations
\begin{eqnarray}\label{ems}
\alpha_{\mathrm{msu}} &=& \frac{4 \gamma^{3}\pm x_{\mathrm{ms}} Y(x_\mathrm{ms},\beta)}{\left(3x_{\mathrm{ms}}-4\beta\right)}\, ,\\
E_\mathrm{msu} &=& \pm\frac{Y(x_\mathrm{ms},\beta)}{\sqrt{x_\mathrm{ms}\left(3x_\mathrm{ms}-4\beta)\right)}}\\ \label{lms}
L_\mathrm{msu} &=&  M\frac{3x_{\mathrm{ms}}\beta-2x_{\mathrm{ms}}^2\pm4\gamma^{3}Y(x_\mathrm{ms},\beta)}{\sqrt{x}(3x_{\mathrm{ms}}^2-4x_{\mathrm{ms}}\beta)^{3/2}}\, ,
\end{eqnarray} 
where the upper sing is valid for $\alpha\geq\alpha_M(\beta)$ and the lower sign is valid for the $\alpha<\alpha_M(\beta)\, .$

The change of sing in $E_{\mathrm{msu}}$ is governed by the following characteristic values of the spin parameter:
\begin{equation}
\alpha_{\mathrm{M}}(\beta)= \frac{4 \left(1 - \beta + \sqrt{1 - 5\beta + 4\beta^2}\right)^{3/2}}{3\sqrt{3}\left(1 - 2\beta + \sqrt{1 - 5\beta + 4\beta^2}\right)}\, ,
\end{equation}  
relevant for the corotating disk, the extreme BH state $\alpha_\mathrm{h}=\sqrt{1-\beta}$ , and also by the existence of mining instability region, see Fig. \ref{am} for more details. We have to state that occurrence of the event horizon causes discontinuous change of the sign of the energy of the marginally stable corotating cicular geodesic. This is standard behavior for the Kerr naked singularity spacetimes and was discussed at length in \cite{Stu:1980:BAC:}.  

In the case of $\beta=0\, ,$ the value 
\begin{equation}
\alpha_\mathrm{M}(\beta=0) = \frac{4}{3}\sqrt{\frac{2}{3}} \sim 1.08866\, .
 \end{equation}
This corresponds to Kerr case studied in \cite{Stu:1981:BULAI:}. Dependence  of $\alpha_{\mathrm{M}}$ on the parameter $\beta$ is depicted in Fig. \ref{am}. 

If the accretion allows approach to the extreme K-N black hole state, there is another subtle point that must be taken into account, being related to the covariant energy of the matter in accretion process. 
 
We can see that for values $0.25<\beta<1$, the energy $E_{\mathrm{ms}}$ is not always well defined - it could approach stable photon orbit as the accreting matter extracts energy of the source -- accretion processes is formally unlimited as covariant energy of matter in accretion could decrease to minus infinity (we have \uvozovky{unlimited mining}). This is the core idea behind mining instability phenomenon - there are no classical marginally stable orbits and the accretion disk formally exists (and extracts energy) down to the stable photon orbit, as state with infinite negative energy is approached. Correspondingly, the energy $E_{\mathrm{ms}}$ decreases (at least in principle) to minus infinity. The mining K-N spacetime becomes unstable because of such energetic particles that no longer can be considered as test particles. Hence the name mining instability. 

For values $\beta>1$ the energy $E_{\mathrm{ms}}$ is well defined and always positive. In the figure we can also see that negative values of $E_{\mathrm{ms}}$ are existing only between $\alpha_{\mathrm{M}}$ and $\alpha_{\mathrm{h}}$. These two functions of parameter $\beta$ seem to approach each other as $\beta$ becomes more negative, but they actually coincide at $\beta=-\infty\, .$
 
\section{Master equation and evolution due to the regime of upper family orbits}

In this section we develop explicit form of the master equation governing mass $M$ evolution in case of the upper family orbits.

By putting Eqs. (\ref{ems}) - (\ref{lms}) into the master equation (\ref{dr}), we can recast the result and find the evolution of mass $M$ with respect to position of marginally stable orbit $x_{\mathrm{ms}}$ in the form:
\begin{equation}\label{eM}
\frac{1}{M}\frac{dM}{dx_{\mathrm{ms}}} = Z_{\pm}(x_{\mathrm{ms}})\, ,
\end{equation}
where 
\begin{eqnarray}
\frac{1}{Z_{\pm}(x)} &=& \frac{2Y^2\left[2x(1-2\beta)-Y^2\right]}{9 (x-1)^2 (x - 2\beta)}-\nonumber \\
&&\frac{x^2(2+\beta)+3x\beta \pm 6\beta (2x-1)\gamma Y}{9 (x-1)^2 (x - 2\beta)}\, .
\end{eqnarray} 
Here the plus sign corresponds to $a_{\mathrm{ms+}}$ case and the minus sign to $a_{\mathrm{ms-}}$. 

Formal solution to the master evolution equation (\ref{eM}) can be expressed via integration
\begin{equation}\label{eMI}
M_{\pm}(x_{\mathrm{ms}}) = M_\mathrm{i\pm}\exp\left\{\int\limits^{x_{\mathrm{ms}}}_{x_{\mathrm{msi\pm}}} Z(x)\, dx \right\}\, ,
\end{equation}
where $M_\mathrm{i\pm}$ denote the initial mass of the central body related to the corresponding boundary conditions, i.e. $x_{\mathrm{msi\pm}}$.  

The formal solution (\ref{eMI}) is well suited to be solved numerically. Due to presence of exponential function in the integral we can immediately see that the evolution of $x_{\mathrm{ms}}$ is governed by the sign of $dM$ in Eq. (\ref{mem}) and subsequently that this sign is solely determined by the sign of the energy $E_{\mathrm{ms}}\, $, i.e. via Eq. (\ref{ems}). We have to stress that in solution procedure of the evolution master equation we can directly use integral form of Eq. (\ref{eMI}) only it the first regime of the evolution of the tidal charge, given by condition $\beta=$ const. In the second regime $b=$ const, we have to solve the differential form of the master Eq. (\ref{eM}) with $\beta = b/M^2\, .$   

\subsection{Singularities of Evolutionary equation}

\begin{figure}[t]
			\begin{center}
			\includegraphics[width=\linewidth]{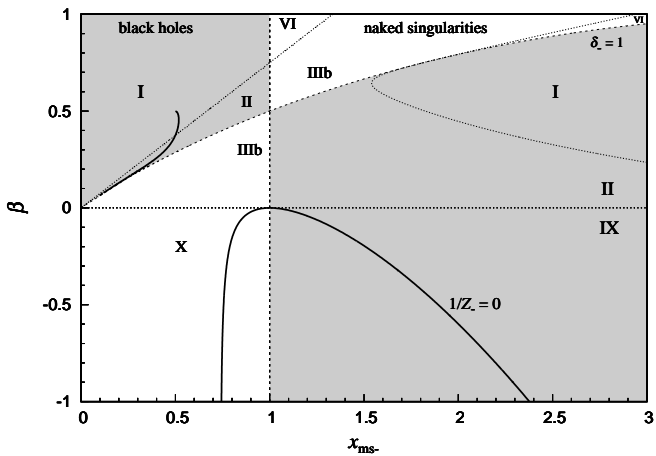}
			\end{center}		
	\caption{\label{s1} Zero points of function $1/Z_{\pm}\, .$} 
\end{figure}

Zero points of function $1/Z_{\pm}$ are shown in the $\beta-x_\mathrm{ms}$ space in Fig. \ref{s1}.

For $0<\beta<1$, or on the region $x<\beta$ for $\beta>1$, the function $1/Z_+$ behaves fully regularly in the regions of interest as its zero points are hidden under the horizon, or in the region of mining instability.

For $1/Z_-$ the singular points where it goes to zero are located in available regions of the space $\beta-x_\mathrm{ms}\, .$ These singular points correspond to the states we call \uvozovky{frozen} for the corrotating Keplerian accretion, as the K-N naked singularity stops its evolution in such a state. Of course, different kind of accretion (e.g. counter-rotating Keplerian) can cause crossing of such a state.  

\section{Evolutionary regime: $\beta=\mathrm{const}$}

We first study evolution in the limiting first regime assuming $\beta =$ const that enables to use simply the integral form of the master evolutionary equation (\ref{eMI}). 

\subsection{Tidal Charge $\beta=0$ (Kerr)}

\begin{figure}[ht]
	\begin{center}
		\includegraphics[width=\linewidth]{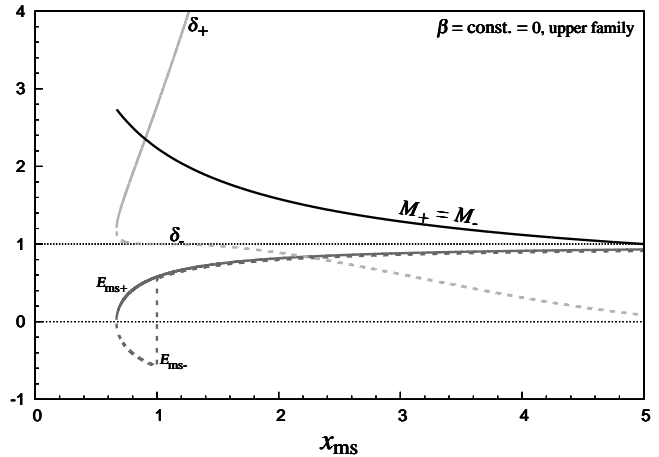}
	\end{center}
	\caption{\label{o1} Evolution of the central mass $M(x)$, specific energy on marginally stable orbit $E_{\mathrm{ms}}$ and auxiliary parameter $\delta_{\pm}\equiv \alpha^2_{\mathrm{ms\pm}}+\beta$  in case of $\beta=0$ corresponding to the Kerr naked singularity and initial condition $M_\mathrm{i}(5)=1\, .$ } 
\end{figure}

For $\beta=0$ we get an analytical solution of the master equation in the form 
\begin{equation}
M_{\pm}(x_{\mathrm{ms}}) = \frac{M_\mathrm{i\pm}}{\sqrt{x_{\mathrm{ms}}}}\, .
\end{equation}
This solution has the same form for both signs and was applied in \cite{Stu:1981:BULAI:} where detailed discussion can be found. An illustrative example is depicted in the Fig. \ref{o1}, where we have set initial values $M(5)=1\, .$  Here we have also depicted energy $E_{\mathrm{ms}}$ and the special parameter $\delta_{\pm} \equiv \alpha^2_{\mathrm{ms\pm}}+\beta$ governing character of the spacetime  with zero tidal charge in this specific case. This parameter distinguishes NS from BH spacetimes, because NS spacetimes are defined by the condition $\alpha^2+\beta>1\, .$ Note that the extremal black holes are defined by the condition $\delta_{\pm} = 1\, .$

In the picture we demonstrate the evolution of mass $M$ in dependence on the parameter $x_{\mathrm{ms}}$. We have two choices of starting setup: either BH or NS spacetime. Both cases have same initial mass $M(5)=1$, but differ in values of the spin parameter, which reflects on different values of $\delta$ parameter.  

The elapsed time of central object evolution with initial values $M_\mathrm{i}$ and $\alpha_\mathrm{msi}$ can be found by integrating the master equation (\ref{dr}) expressed in the form
\begin{equation}
\frac{d M}{d t} = E_{\mathrm{ms}}\frac{d\mu}{dt}\, ,
\end{equation} 
assuming that the accretion rate relative to observers at infinity $d\mu/dt$ is known. This accretion rate can be approximated, e.g, by the observed velocity of the matter density leaving marginally stable orbit - this is non zero even for density of test particles. In case of ordinary Kerr superspinars this was done in \cite{Stu:Hle:Tru:2011:}. 

Here we are not considering the time evolution and the amount of matter accreted on the central object, postponing these studies for following papers. 

First, let us consider evolution of NS spacetime. Parameter $\delta$ must be larger that 1, therefore at the starting point $x_{\mathrm{ms}}=5$ we see from the Fig. \ref{o1}, that system is governed by the + sign quantities. The actual value of $\delta_+=32.1358$ cannot be read from the picture itself. 

Since the corresponding $E_{\mathrm{ms+}}$  is positive, we can conclude that the mass $M$ of the NS increases due to the accretion. Therefore, the radius of the marginally stable orbit $x_{\mathrm{ms}}$ and its energy $E_{\mathrm{ms}}$ decrease due to accretion and also the spin of the naked singularity decreases. The evolution in the Fig. \ref{o1} is translated into the movement of the systems parameters to the left. In this picture, we do not see the speed of the evolution, but just its general behavior, which is governed only by the sign of $E_{\mathrm{ms}}$.

The system evolves into the most left point where the energy is zero and the two branches $(\pm)$ of the solution switch places (in this case the scenario is simpler because $M_+\equiv M_-$). System is still a NS spacetime, but now the energy of the marginally stable orbit is negative and so the central mass $M$ decreases. The energy of the marginally stable orbit and the naked singularity spin are decreasing, but the radius $x_{\mathrm{ms}}$ is now increasing! In Fig. \ref{o1} we are now shifting back, i.e. to the right. 

The system continuously evolves to the point ($x_{\mathrm{ms}}=\delta_{-}=1$) where it collapses into the extremal BH spacetime. As horizon of extremal BH emerges, the energy of marginally stable orbit $E_{\mathrm{ms-}}$ discontinuously jumps to positive values (see Fig. \ref{am} and detailed discussions in \cite{Stu:1980:BAC:,Stu:1981:BULAI:}). Therefore evolution has to again switch its direction and the BH-accretion disk system remains formally the extremal BH spacetime. Evolution, at least in terms of dimensionless quantities is stopped. As discussed in \cite{Stu:1980:BAC:,Stu:1981:BULAI:,Stu:Hle:Tru:2011:}, the real transcendence to the BH state could lead to a non-extreme BH state due to an instability of the Keplerian disk in the region up to the $E_\mathrm{ms}$ of the BH state. 

\subsection{Tidal Charge $0<\beta<0.25$}

\begin{figure*}
	\begin{minipage}{.5\linewidth}
		\begin{center}
			\includegraphics[width=1\linewidth]{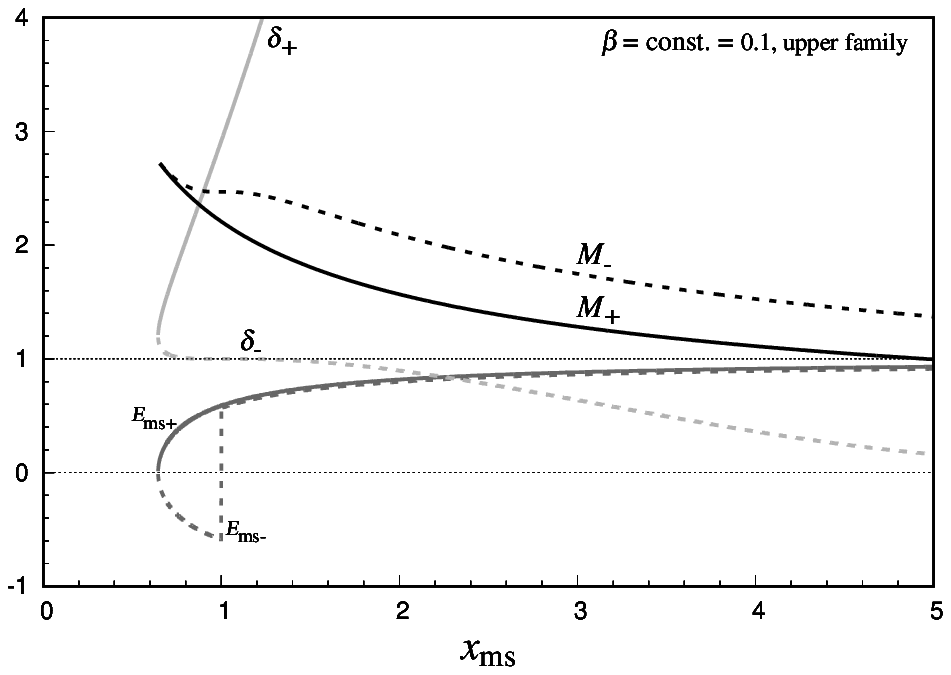}
		\end{center}
	\end{minipage}\hfill
	\begin{minipage}{.5\linewidth}
		\begin{center}
			\includegraphics[width=1\linewidth]{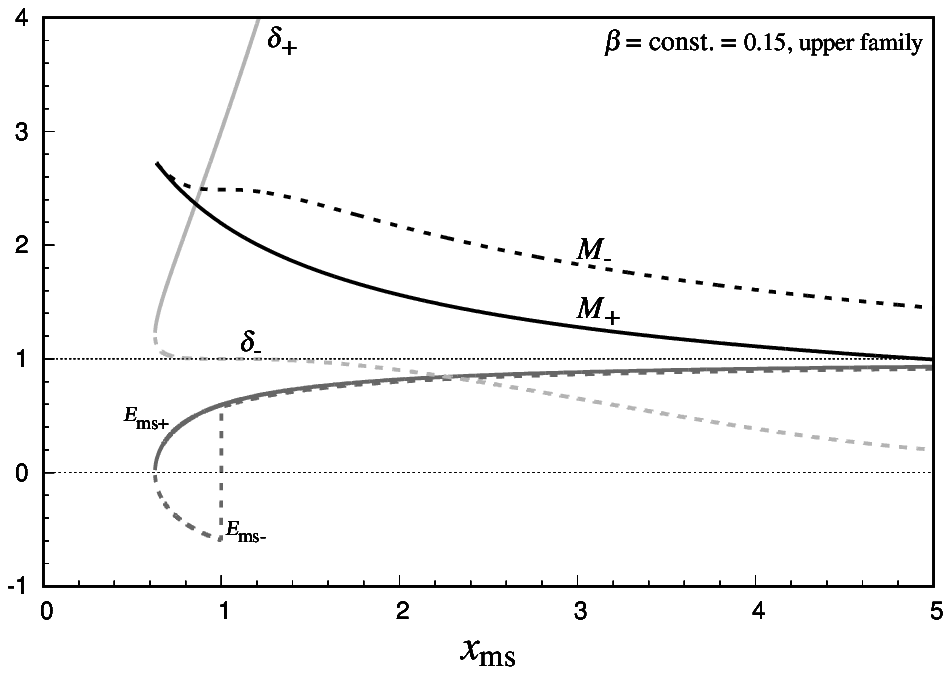}
		\end{center}
	\end{minipage}
	\begin{minipage}{.5\linewidth}
		\begin{center}
			\includegraphics[width=1\linewidth]{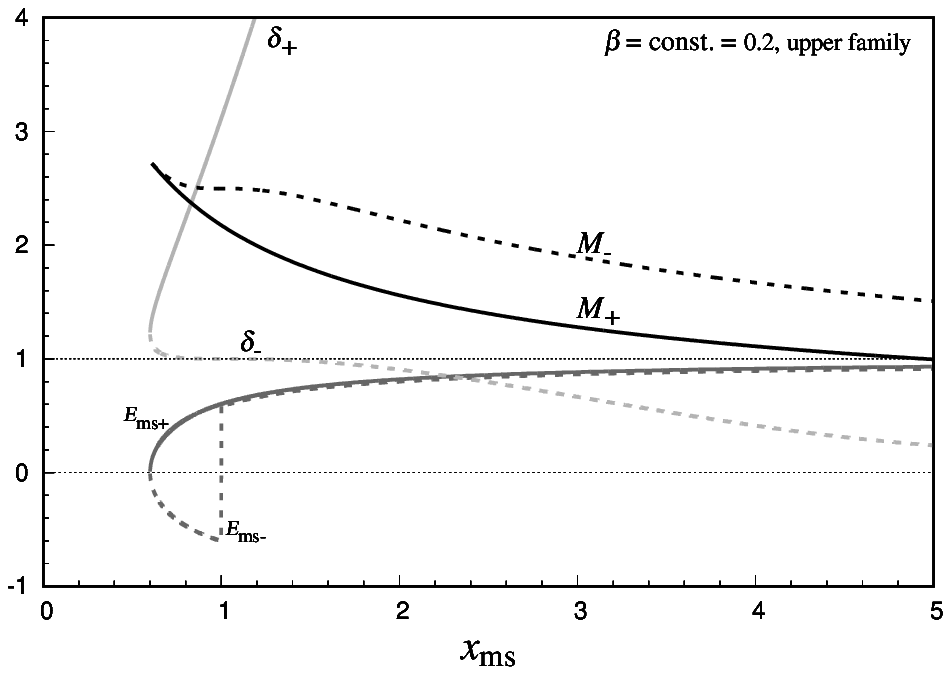}
		\end{center}
	\end{minipage}\hfill
	\begin{minipage}{.5\linewidth}
		\begin{center}
			\includegraphics[width=1\linewidth]{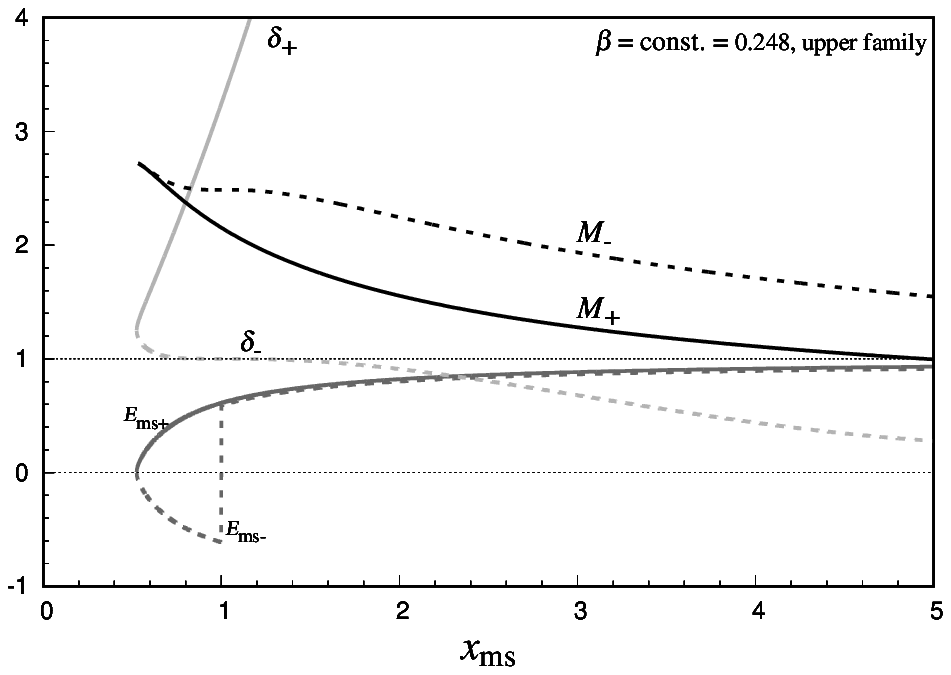}
		\end{center}
	\end{minipage}
	\caption{\label{o2} Evolution of the central mass $M(x)$, specific energy on marginally stable orbit $E_{\mathrm{ms}}$ and auxiliary parameter $\delta_{\pm}$. Here, there are four examples of values $\beta \in \left\{0.1;0.15;0.2;0.248\right\}$ with the same initial condition $M_\mathrm{i}(5)=1\, .$} 
\end{figure*}

The case of constant tidal charge parameter $\beta$ spanning $0<\beta<0.25$ is essentially same as previous one with $\beta=0\, .$ The difference is that two solutions $M_+$ and $M_{-}$ are no longer identical, as we can see in the Fig. \ref{o2}. In this figure there are four examples of evolution with $\beta \in \left\{0.1;0.15;0.2;0.248\right\}\, .$ 

The character of the evolution is qualitatively identical to the case of $\beta=0$, therefore, there is a shift in the sign of the energy of the marginally stable circular geodesic corresponding to the shift from the regime of increasing of the central mass to the regime of its decrease. Finally, the extreme BH state is reached, where the jump of the energy of the marginally stable orbit occurs.

\subsection{Tidal charge $0.25<\beta<1$}

\begin{figure}[ht]
\begin{center}
\includegraphics[width=1\linewidth]{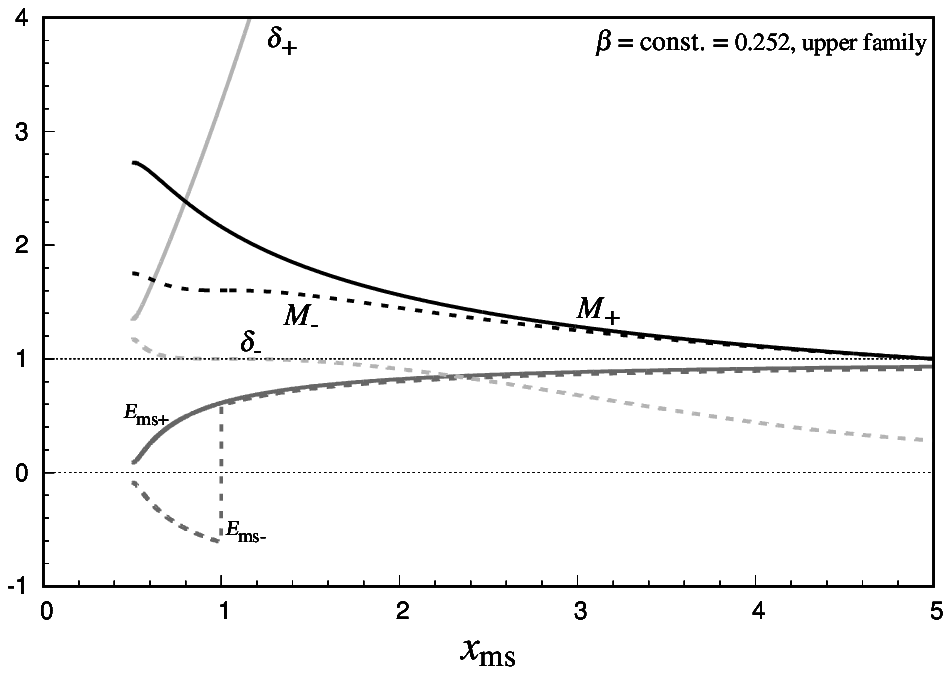}
\end{center}
\caption{\label{me0252} Evolution of the central mass $M(x)$, specific energy on marginally stable orbit $E_{\mathrm{ms}}$ and auxiliary parameter $\delta_{\pm}$ in case of $\beta=0.252$ and initial conditions $M_\mathrm{i}(5)=1\, .$ Graph demonstrates that $E_{\mathrm{ms}+}$ and $E_{\mathrm{ms}-}$ are separated functions in cases $\beta>0.25\, .$} 
\end{figure}

In the case of $0.25<\beta<1$ the situation is fundamentally different. The two branches of energies $E_{\mathrm{msu}}$ (lines on the left in the Fig. \ref{o2}) separate into the two detached functions starting at $\beta=0.25$. This is demonstrated in the Fig. \ref{me0252}.

In this figure, we see that functions $E_{\mathrm{ms+}}$ and $E_{\mathrm{ms-}}$ are separated and the system cannot evolve from $+$ into $-$ case. The evolution is therefore formally stopped at the maximum of $M_+$ where the energy $E_{\mathrm{ms+}}$ is still positive and therefore systems mass should go up. 

This strange behavior is caused by the mining instability (see \cite{Bla:Stu:2016:}). The combination of values of the $\alpha$ and $\beta$ parameters are at this point such that the spacetime belongs to class IIIa. Therefore, the picture is no longer sufficient for describing the Keplerian accretion and enters special class IIIa of K-N naked singularity spacetimes. 

We might now admit that the numerical analysis of the master equation (\ref{dr}) should at this point collapse, because $E_{\mathrm{ms}}$ formally approaches minus infinity. In fact both $E_{\mathrm{ms}}$ and $L_{\mathrm{ms}}$ are for IIIa class spacetimes formally infinitely negatively large, but the fraction of these two entities
\begin{equation}
\frac{E_{\mathrm{ms}}}{L_{\mathrm{ms}}}\equiv \frac{1}{B}
\end{equation}
is well defined, being just the reciprocal of the impact parameter $B$ for photons on the stable circular null geodetics [16]. 

On the other hand in the Fig. \ref{me0252} we can see that evolution of the BH (function $M_-$) with the same initial mass, proceeds normally to the extreme BH, where it again stabilizes. This is so because the mining instability phenomenon occurs only for NS spacetime. 

Note that in the K-N NS spacetimes of class VI, located between the class IIIa of K-N NS spacetimes and the K-N BH spacetimes, the situation is similar to the IIIa NS spacetimes -- the evolution is again finishing, after transition across a "false" marginally stable orbit, in the state with unlimited descending of the energy of the matter in accretion, corresponding to the mining instability requiring a transcendence due to non-linear phenomena as in the IIIa spacetimes.

\subsection{Tidal Charge $\beta>1$}

Some types of NS spacetimes with $\beta>1$ do not allow Keplerian accretion, at least not the simple model of thin disk. It is due to prevalent existence of the innermost stable circular orbit (ISCO) at the radius $x=\beta\, .$ More detailed explanation can be found in \cite{Bla:Stu:2016:}, but essentially, at $x=\beta$ radius the matter is trapped and can never fall onto the singularity. This strange behavior suggests to consider other transcendental models of the NS - accretion disk systems, as we have evolution where the NS is fixed and due to the accretion an additional compact structure grows around $x=\beta\, ,$ giving eventual final state as black Saturn \cite{Elv:Fig:2007:}. 

This mechanism is different from the mining instability and we are not going to survey it in more details. Moreover, there are studies which restrict the possible values of the tidal charge $\beta$ due to the classical tests in Solar system \cite{Boh-etal:2008:CLAQG:} and physics around compact objects \cite{Kot-Stu-Tor:2008:CLAQG:}. From these findings it is clear that if tidal charge exists, its absolute value has to be quite small and certainly not bigger than 1. Therefore, from the astrophysical point of view the cases $|\beta|>1$ are not relevant.  

\subsection{Tidal charge $\beta<0$ }

\begin{figure}[t]
	\begin{center}
		\includegraphics[width=1\linewidth]{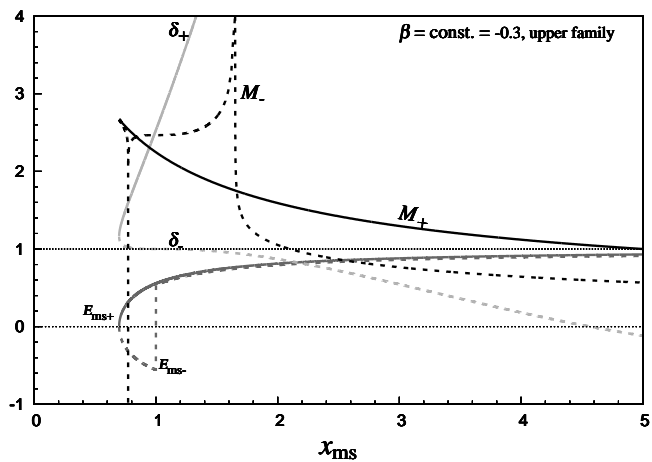}
	\end{center}
	\caption{\label{mem03} Evolution of the central mass $M(x)$, specific energy on marginally stable orbit $E_{\mathrm{ms}}$ and auxiliary parameter $\delta_{\pm}$ in case of $\beta=-0.3$ and initial condition $M_\mathrm{i}(5)=1 \, .$ which is equivalent to $\alpha_\mathrm{i} = $. The part between two singular points lying at $x=0.77153$ and $x=1.650734$ has to be properly understand as general behaviour, because the curve cannot communicate with the initial data. }
\end{figure}

The spacetimes with negative values of tidal charge (types VIII and IX) behave pathologically for naked singularities. In the Fig. \ref{mem03} we can see how. There are two singular points in solution $M_-$. Positive infinity can be physically interpreted as a stable \uvozovky{frozen} BH configuration. The accretion onto this BH state just increases mass $M_-$ and dimensionless quantities $\beta\, \alpha$ and $x_{\mathrm{ms}}$ remain unchanged. 

Negative infinity point is hard to interpret meaningfully. Perhaps we can say that the spacetime at this point is fundamentally unstable towards accretion and the solution is invalid. Fig. \ref{mem03} suggests that the mass of NS spacetime $M_-$ is due to the accretion process reduced to zero! Such destruction of NS due to simple accretion is hard to take seriously. 

It should be stressed that both these strange states of the evolutionary singular behavior are not stable characteristics of these K-N NS spacetimes as any violation of the Keplerian corotating accretion enables crossing of these states.  

\section{Evolution due to counter-rotating  Keplerian disk}

The situations containing the configuration with central object and counter-rotating Keplerian disk (lower family) are quite easy to analyze due to the fact that counter-rotating marginally stable orbits are positioned much farther from the central object and there are no demonstration of extraordinary phenomena related to this family of circular orbits. Therefore, the behavior of such configurations is well defined, the energy of marginally stable orbits is always positive and there is no jump in crossing between the NS and BH states. The spacetime evolution due to the counter-rotating Keplerian accretion is determined by the relations\begin{eqnarray}\label{ems-c-r}
\alpha_{\mathrm{msu}} &=& \frac{-4 \gamma^{3} + x_{\mathrm{ms}} Y(x_\mathrm{ms},\beta)}{\left(3x_{\mathrm{ms}}-4\beta\right)}\, ,\\
E_\mathrm{msu} &=& \frac{Y(x_\mathrm{ms},\beta)}{\sqrt{x_\mathrm{ms}\left(3x_\mathrm{ms}-4\beta)\right)}}\\ \label{lms2}
L_\mathrm{msu} &=&  -M\frac{3x_{\mathrm{ms}}\beta-2x_{\mathrm{ms}}^2 + 4\gamma^{3}Y(x_\mathrm{ms},\beta)}{\sqrt{x}(3x_{\mathrm{ms}}^2-4x_{\mathrm{ms}}\beta)^{3/2}}\, ,
\end{eqnarray} 

During the evolution the NS smoothly converts into the BH ($\delta_- =1$). Crucial is the smooth conversion of NS into BH which is quite regular in this case, with no discontinuities and arising of unstable regions of the Keplerian disk that has to be swallowed into the BH as in the case of corotating Keplerian disk (see discussions in \cite{Stu:1981:BULAI:,Stu:Hle:Tru:2011:}). Since we are in the counter-rotating case, the $E_{\mathrm{ms-}}$ is positive. The evolution continues until the point with the spin $\alpha=0$ (minimum of $\delta_-$). At this point the accreting matter starts to spin-up the central object in the opposite direction. This means that now we are dealing with the upper family (corotating) evolution and from the picture we see that system will eventually evolve into a stable extremal BH.     

\begin{figure}[t]
\begin{center}
\includegraphics[width=\linewidth]{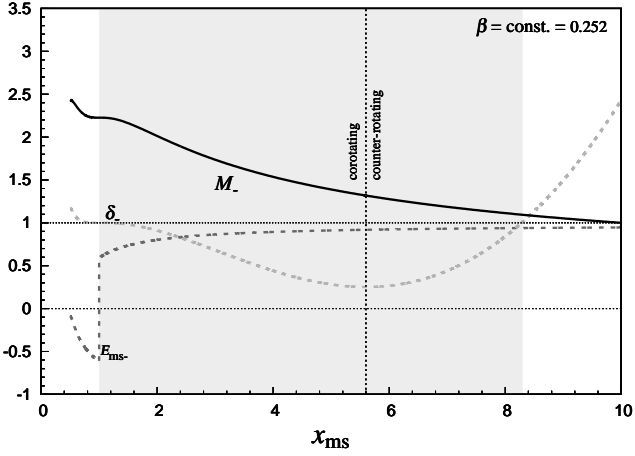}
\end{center}
\caption{\label{coun} Evolution of the central mass $M(x)$, specific energy on marginally stable orbit $E_{\mathrm{-}}$ and auxiliary parameter $\delta_{-}$ in case of $\beta=0.252$ and initial condition $M_\mathrm{i}(10)=1\, .$ Counter-rotating (shaded area) case together with corotating (white area) case. Here we have depicted only minus branch functions to simplify the picture. The $M_{+}$ function would behave similarly as $M_-$, but instead of end up in extreme black hole state it would evolve into the mining instability.}
\end{figure}

As an example see Fig. \ref{coun}. Here we have depicted evolution of a NS with mass $M_\mathrm{i}(10)=1\, .$ The energy of marginally stable counter-rotating orbits $E_{\mathrm{ms}}$ are always positive (see shaded part of Fig. \ref{coun}). Therefore, the mass of the system increases, while the spin $\alpha$ decreases, which can be seen in the evolution of parameter $\delta_-$. 

We have to stress that the regular character of the counter-rotating Keplerian accretion holds also in the case of the mining K-N naked singularities (classes IIIa, VI). Therefore, their conversion into extreme K-N BH is possible in this way.   
 
\section{Evolutionary regime: $b(\equiv Q^2) = \mathrm{const}$}

\begin{figure}[ht]
\begin{center}
\includegraphics[width=\linewidth]{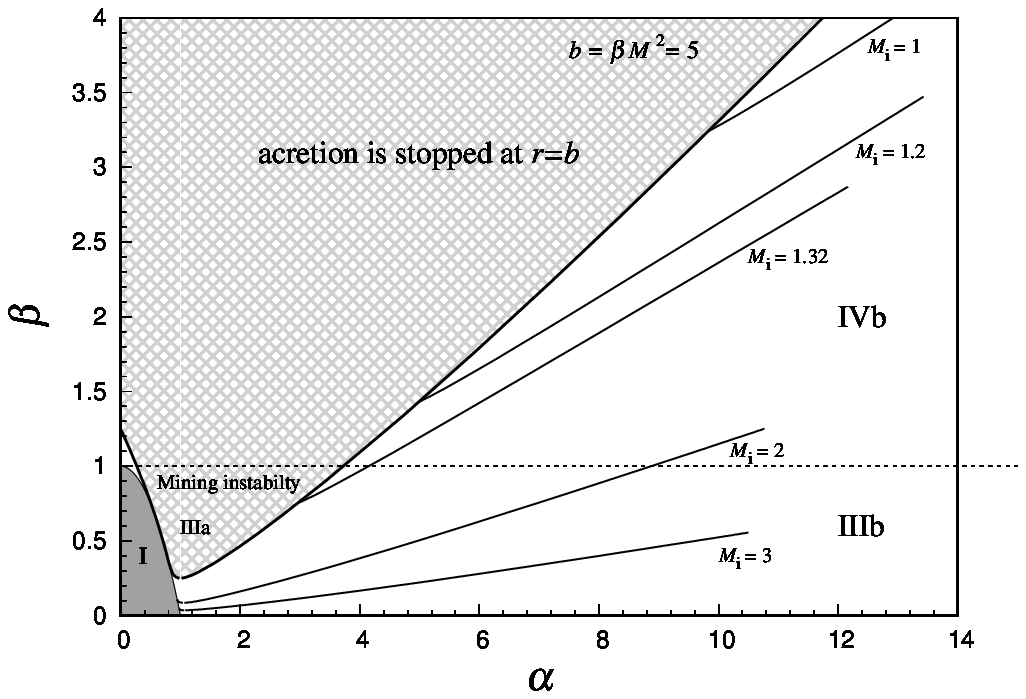}
\end{center}
\caption{\label{bv} Evolution of spacetime in $\alpha - \beta$ parameter space due to accretion. Tidal charge is held constant ($b=5$), therefore $\beta$ is changing.}
\end{figure}

\begin{figure}[ht]
\begin{center}
\includegraphics[width=\linewidth]{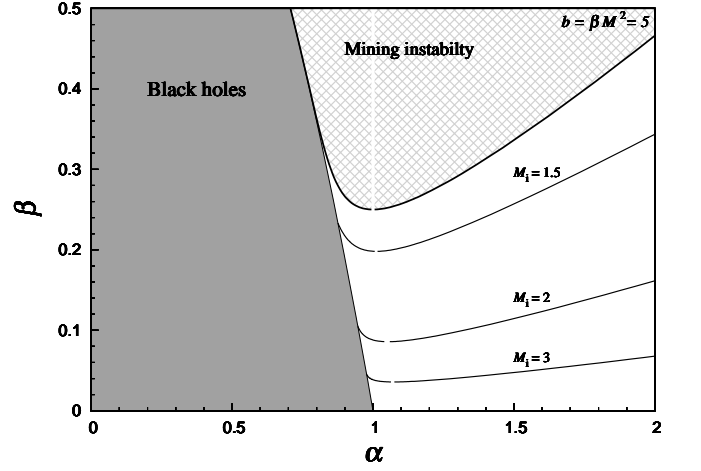}
\end{center}
\caption{\label{bm} Evolution of spacetime in $\alpha - \beta$ parameter space due to accretion.  Tidal charge is held constant ($b=5M^2$), therefore $\beta$ is changing.}
\end{figure}

In this section we study the evolution of the central object during corotating Keplerian accretion assuming tidal charge $b\equiv M^2\beta$ being constant. For positive values of tidal charge parameter ($b>0$) this case corresponds to evolution of K-N spacetime where $b=Q^2\, $ and the matter in accretion disk is charge neutral.

The master equation for second case can be obtained simply by the substitution
\begin{equation}
\beta \rightarrow \frac{b}{M^2}\, .
\end{equation}     

Although the master equation looks quite complicated and cannot be expressed in the form of an integral, but has to be solved numerically, its essential behaviour is very similar to previous case ($\beta=\mathrm{const.}$). The main difference is that the dimensionless parameter $\beta$ is changing due to change of the mass $M$. In Fig. \ref{bv} we have taken into account this fact and shown several evolutions of the central object with different initial masses $M_\mathrm{i}$. We can see that evolution is also stopped if the mining instability region is reached where we have to consider more sophisticated model where simple thin Keplerian accretion disk has to be finally modified by non-linear phenomena governing interaction of the NS and the disk.  

The evolution is also stopped in K-N naked singularity classes with tidal charge parameter $b>M^2$ where the matter cannot fall onto the NS being gathered at $r=b/M$, where innermost (but not marginally) stable circular orbit is located. In these cases we also have to assume a transcendence of the system NS + corotating keplerian disk into some new object due to non-linear general relativistic phenomena.  

Therefore, there are three possible scenarios:  

\begin{enumerate}

\item the evolution of the central object goes in a way that the path depicted in $\alpha-\beta$ parameter space misses any problematic classes of naked singularity spacetimes and then the central object evolves into an extremal BH. This is clearly illustrated in Fig. \ref{bm},

\item the evolution goes into regions with $\beta>1$. The accretion process is then stopped at $r=b/M$ orbit. In this case we can consider modified evolution due to Keplerian accretion, related to evolution of the Komar mass reflecting sum of masses and spins of the central object and torodial structure around $r=b/M$ created by accretion. Then the central object could even become a black Saturn. The similar results were found for  Kehagias--Sfetsos NS in \cite{Stu:Pug:Sche:Kuc:2015:},
   
\item the evolution is stopped at border of class IIIa region - the mining instability spacetime. This situation is even more problematic than the $b>M^2$ case, because the mining instability demonstrates extraordinary properties. Even one particle falling onto the NS can obtain large amount of energy and become too heavy to be considered as a test particle. The K-N solution is in this case clearly demanding transcendence due to the physical process of corotating accretion.     

In the regime of counter-rotating Keplerian accretion the evolution to the extreme K-N black hole state is possible from any K-N naked singularity state with $b<M^2\, .$

\end{enumerate} 

\begin{figure}[t]
	\begin{center}
		\includegraphics[width=\linewidth]{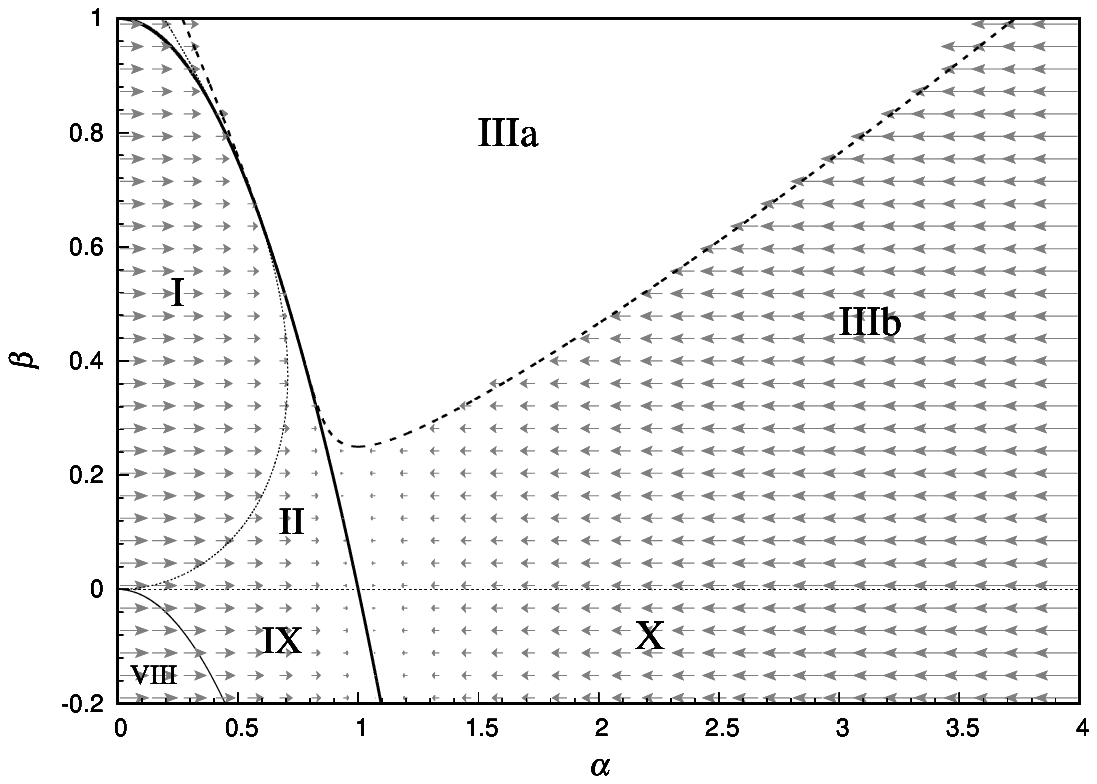}
	\end{center}
	\caption{\label{df11} Directional field for the $b=\,$const model.}
\end{figure}

\begin{figure}[t]
	\begin{center}
		\includegraphics[width=\linewidth]{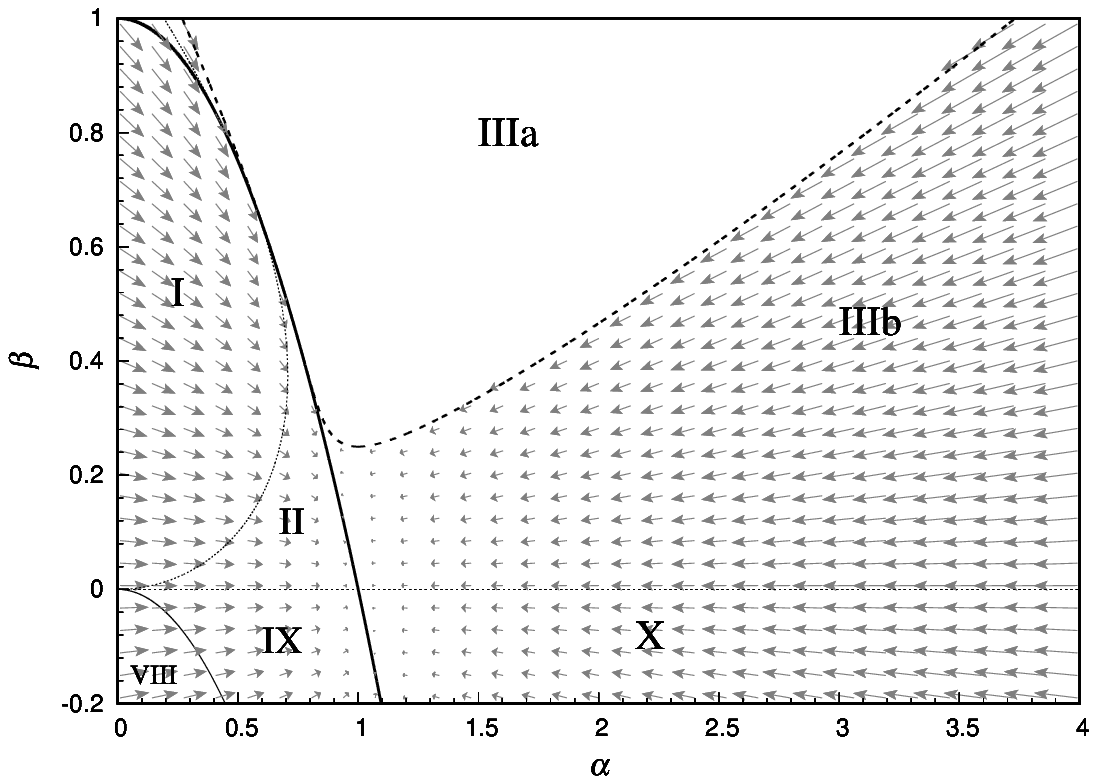}
	\end{center}
	\caption{\label{df1} Directional field for the $b=\,$const model.}
\end{figure}

\begin{figure}[t]
	\begin{center}
		\includegraphics[width=\linewidth]{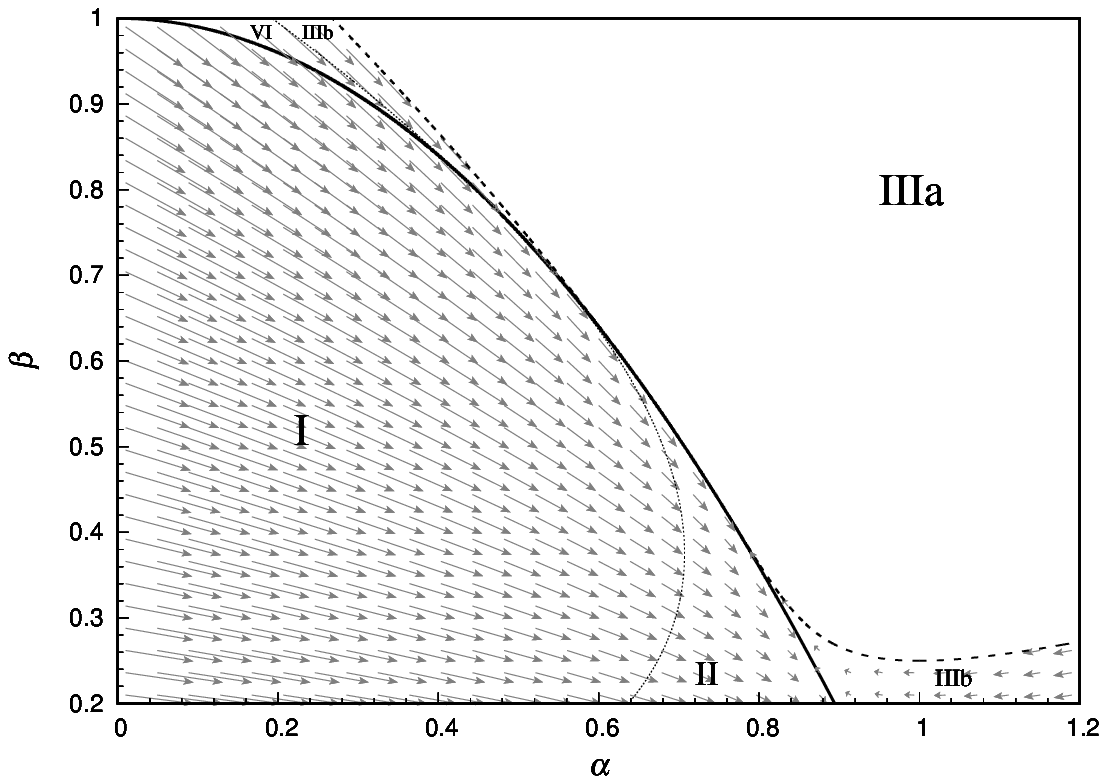}
	\end{center}
	\caption{\label{df2} Directional field for the $b=\,$const model.}
\end{figure}  

\section{Directional fields}

To illustrate behaviour of the solutions of the master equation we can express the directional field in the $\alpha-\beta$ parameter space. Combining master equation (\ref{dr}) and Eq. (\ref{mem}) we find that 
\begin{equation}
d\alpha = \left(\frac{L_{\mathrm{ms}}}{M}-2\alpha_{\mathrm{ms}} E_{\mathrm{ms}}\right)\frac{d\mu}{M}\, .
\end{equation}
The differential of $\beta$ is zero in the first regime and it is given by 
\begin{equation}
d\beta = -2\beta E_{\mathrm{ms}}\frac{d\mu}{M}\, ,
\end{equation}
in the second regime of tidal charge evolution (see section \ref{models}). Combining these two equations we arrive to the expression giving the directional field
\begin{equation}
\frac{d\alpha}{d\beta} = \frac{\alpha_\mathrm{ms}}{\beta} - \frac{L_\mathrm{ms}}{2\beta M E_\mathrm{ms}}\, .
\end{equation}

Since $L_{\mathrm{ms}}/M$ does not depend on the mass $M$, we can create simple picture of directional field, representing local behaviour of the master equation solution.  In Fig. \ref{df11} we have depicted examples of directional field for the first regime. This example is trivial with respect to $\beta$ parameter since this parameter does not change during accretion. In the picture, we can see that K-N black hole classes (I,II,VIII,IX) always evolve to right, towards the extremal black hole state. Some classes which represent K-N naked singularity spacetimes evolve to the left, till they reach extremal black hole state or mining unstable class IIIa. 

Similar situation can be seen in the Fig. \ref{df1} where we have depicted practically the same example, but this time under the second regime (where parameter $b$ is fixed during accretion). Here we find similar behaviour and in the detailed Fig. \ref{df2} we also can see that some naked singularity classes (VI and part of IIIb) evolve to the right till they become extremal black hole or potentially reach IIIa class. For completeness we also demonstrate the directions of the BHs (classes I,II,VII and IX) that always tend to the extreme BH states.

\section{Conclusions} 

In the present paper we concern attention on the evolutionary ways and possible final states of the K-N naked singularities reflected by the dependence of the radius of marginally stable orbit, if such orbit exist allowing fall of matter onto the centre of the naked singularity spacetime. 

Conversion of a K-N naked singularity to an extreme K-N black hole is possible in the standard way in linear regime for counter-rotating Keplerian accretion, for corotating accretion can imply a transcendence to non-linear regime of evolution in two regimes.

In fact, the standard ISCO serving as a marginally stable orbit from which matter can freely fall onto the singularity exists only in a limited class of the braneworld K-N naked singularities (IIIb and X) enabling the standard evolutionary regime as known from the case of Kerr naked singularities \cite{Stu:1981:BULAI:,Stu:Hle:Tru:2011:}. There are however K-N naked singularities with ISCO at $x=\beta$ that are not marginally stable but they are related to final stable state of the accretion process where accreting matter is accumulated forming a torodial structure of total covariant energy $E_\mathrm{t}= E_\mathrm{ISCO}(x=\beta)\Delta \mu_\mathrm{t}$, where $\Delta \mu_\mathrm{t}$ denotes the total amount of accreted rest mass. From the point of view of the distant observer, the evolutionary way could be given by the sum of $M+E_\mathrm{t}\, ,$ and $J+L_\mathrm{t}$ where $L_\mathrm{t}$ is total axial angular momentum of the accreted matter. Completely different is the character of the \uvozovky{transcendental} mining unstable class IIIa of K-N naked singularities.   

We demonstrated that mining instability of K-N singularity spacetime represents a challenge to most simple cases of Keplerian accretion. And this can be understood within the framework of braneworld scenarios having tidal charge $0.25<b<1$ or within the classical framework of K-N singularity having $0.25<Q^2<1\, .$   

Cases with $\beta>1$ have to be studied more carefully, because matter is gathering at $r_{\mathrm{ISCO}}=b\, .$

In the case of BH spacetimes with negative values of the tidal charge parameter we demonstrated another kind of possible strange behavior of the corotating Keplerian accretion -- namely the stable configuration of BH with respect to accretion during which the mass $M$ of the BH grows, but all dimensionless quantities describing the spacetime $(\beta,\alpha,x)$ remain fixed during the evolution. This behavior is not so unexpected, because it was showed that negative values of $b$ strengthened the gravitation field and the spacetime behaves very similarly like BH with greater $M$. 
   
\section*{Acknowledgements}

The authors acknowledge the institutional support of the Research Centre for Theoretical Physics and Astrophysics, Institute of Physics, Silesian University in Opava. ZS acknowledges the Czech Science Foundation grant No. 19-03950S and the internal student grant No. SGS/12/2019 of Silesian University in Opava.

\end{document}